\documentclass[floatfix,aps,twocolumn,preprintnumbers,amsmath,amssymb,nofootinbib,superscriptaddress,showkeys,showpacs]{revtex4}

\usepackage{graphicx,color}
\usepackage{amsmath,amssymb,bm}
\usepackage{dcolumn}

\newcommand{\bea}{\begin{eqnarray}}
\newcommand{\eea}{\end{eqnarray}}
\newcommand{\be}{\begin{equation}}
\newcommand{\ee}{\end{equation}}
\newcommand{\np}{{\bf p}}

\newcommand{\nh}{{\bf h}}

\newcommand{\nq}{{\bf q}}

\newcommand{\Qbar}{\not{\!Q}}

\newcommand{\kbar}{\not{\!k}}
\newcommand{\Pbar}{\not{\!P}}
\newcommand{\tauvec}{\mbox{\boldmath $\tau$}}

\def\XXint#1#2#3{{\setbox0=\hbox{$#1{#2#3}{\int}$}
     \vcenter{\hbox{$#2#3$}}\kern-.5\wd0}}

\def\1{\'{\i}}

\begin{document}

\title{Meson-exchange currents and
  superscaling analysis with relativistic effective mass
  of quasielastic electron scattering from $^{12}$C 
}
\author{V.L. Martinez-Consentino}\email{victormc@ugr.es} 
  \affiliation{Departamento de
  F\'{\i}sica At\'omica, Molecular y Nuclear \\ and Instituto Carlos I
  de F{\'\i}sica Te\'orica y Computacional \\ Universidad de Granada,
  E-18071 Granada, Spain.}

\author{I. Ruiz Simo}\email{ruizsig@ugr.es} \affiliation{Departamento de
  F\'{\i}sica At\'omica, Molecular y Nuclear \\ and Instituto Carlos I
  de F{\'\i}sica Te\'orica y Computacional \\ Universidad de Granada,
  E-18071 Granada, Spain.}

\author{J.E. Amaro}\email{amaro@ugr.es} \affiliation{Departamento de
  F\'{\i}sica At\'omica, Molecular y Nuclear \\ and Instituto Carlos I
  de F{\'\i}sica Te\'orica y Computacional \\ Universidad de Granada,
  E-18071 Granada, Spain.}

\date{\today}

\begin{abstract}
\rule{0ex}{3ex} 

We reanalyze the scaling properties of inclusive quasielastic electron
scattering from $^{12}$C by subtracting from the data the effects of
two-particle emission.  A model of relativistic meson-exchange
currents (MEC) is employed within the mean field theory of nuclear
matter, with scalar and vector potentials that induce an effective
mass and a vector energy to the nucleons.  A new phenomenological
quasielastic scaling function is extracted from a selection of the
data after the subtraction of the 2p-2h contribution. The resulting
superscaling approach with relativistic effective mass (SuSAM*) can be
used to compute the genuine quasielastic cross section without
contamination of the 2p-2h channel that can then be added separately
to obtain the total quasielastic plus two-nucleon emission response.

\end{abstract}



\maketitle

\section{Introduction}

Inclusive electron scattering provides information about the
quasielastic response of nuclei, which 
is dominated by one-nucleon emission.
The modeling of these reactions is a trending topic
due to their direct application to neutrino experiments 
\cite{Ama20,Mos16,Kat17,Alv14,Ank17,Ben17}. 
Specifically, several quasi-elastic charge-changing (CC) 
experiments with neutrinos and antineutrinos
have been performed (MiniBooNE, MINERvA, T2K, NOMAD,\ldots)
\cite{Nomad09,Agu10,Agu13,Fio13,Abe13,Abe16,Abe18}
for a variety of targets.  
This allows comparisons to be made with the various existing nuclear models 
\cite{Mar09,Nie11,Gal16,Meg16,Meg16b,Meg14,Ank15,Gra13,Pan16,Mar16}.
The differences found between the various models 
imply a non-negligible systematic error
in neutrino oscillation experiments coming from the
difficulty in the theoretical description of the neutrino-nucleus interactions.
Electron scattering reactions allow us
to fix the kinematics and study with precision the differential cross
section in detail, while in neutrino experiments only flux-averages
can be measured.

Many of the nuclear models that have been applied to the $(e,e')$
region \cite{Lov16,Pan15,Ank15b,Roc16}, are based on non-relativistic
nuclear physics. One of the difficulties is to extend these and other
models to the relativistic regime in the kinematics region of
interest, with momentum transfer $q \sim 1$ GeV/c \cite{Ama02,Ama05}.
The simplest fully relativistic model is the relativistic Fermi gas,
that does not includes interactions between nucleons. Beyond that, the
relativistic mean field (RMF) theory allows to include the
relativistic interaction of nucleons with scalar and vector potentials
\cite{Cab07,Udi99,Ama05}. In particular, relativistic dynamics produces
an enhancement of the transverse response \cite{Bod14,Gal16} that goes
in the direction to reproduce the $(e,e')$ data.  Another key
ingredient for the nuclear inclusive cross section is the two-nucleon
emission (2p2h) channel produced by meson-exchange currents (MEC)
\cite{Gil97,Sim16,Nie17}. The effect of this 2p2h contribution with
relativistic dynamics is explored in the present work.

An alternative to the nuclear models are those based on scaling and
superscaling (SuSA) \cite{Alb88,Don99a,Don99b,Mai02,Ama04,Ama05}, where a
phenomenological scaling function is obtained from the experimental
longitudinal response function $R_L(q,\omega)$, by dividing by a single
nucleon averaged cross section and making a change of variable $\omega
\rightarrow \psi(q,\omega)$ 
such that the resulting longitudinal scaling function 
$ f_L(\psi)$
is centered around the interval $(-1,1)$, and 
do not depend much on $q$. An appropriate scaling variable
 is found by using the theory of the
relativistic Fermi gas to map of the $\omega$ interval
$(\omega_{min},\omega_{max})$ into the $\psi$-interval $(-1,1)$,
where $\omega_{min,max}$ are bounds of the RFG response functions for $q$-fixed.
The value $\psi=0$ correspond to the maximum of the QE peak.
The resulting SuSA model uses the phenomenological scaling function $f_L(\psi)$
to construct the cross section by multiplying by the single nucleon factor.
The SuSA initial assumption was that the transverse response is obtained 
with a transverse scaling function $f_T=f_L$. However 
this hypothesis is  not satisfactory to reproduce the data.
Therefore the super-scaling model has been improved into the SuSA-v2
approach, by using the theory of relativistic mean field (RMF) model of finite
nuclei \cite{Cab07,Hor91} to construct the enhanced 
transverse scaling function $f_T$ by a fit  to $(e,e')$ 
cross section data including 2p-2h MEC and inelastic
contributions \cite{Gon14,Meg16b}.

The goal of this work is to present a model that shares and unifies
the ideas of the RMF, superscaling, and MEC in a consistent way.  The
idea is the extend our previous works on superscaling with relativistic
effective mass \cite{Ama15,Ama17,Mar17,Ama18} to include the 2p2h
contribution, taking into account the interactions of nucleons with
the relativistic mean field.  The attractive scalar potential is
accounted for in the relativistic effective mass $m_N^*< m_N$, while
the vector potential produces a repulsive energy, that has an important 
effect in the MEC.  The resulting 2p2h
MEC matrix elements are modified in the medium due to the interaction
with the relativistic scalar and vector potentials.  In this way the
new model SuSAM*+MEC introduced in this work includes dynamical
relativistic effects both in the scaling function $f^*(\psi^*)$ and in
the MEC.  The final goal is to have a consistent model to be applied
in the future to neutrino scattering as in Ref. \cite{Rui18}.

In the original SuSAM* studies \cite{Ama15,Ama17,Mar17,Ama18} a new
super scaling function $f^*(\psi^*)$ was obtained from the electron
scattering cross section data, by using the scaling variable $\psi^*$
of the RMF.  The model can describe a large amount of the quasielastic
electron data for many nuclei within a theoretical error band.  Note
that in the SuSAM* there is only one scaling function because the
relativistic mean field generates the transverse response enhancement
\cite{Ros80,Ser86,Weh93}.

Here we improve the SuSAM* analysis by subtraction of the 2p2h cross
section from the inclusive cross section before extracting the scaling
function.  In this way we avoid to include possible 2p2h contamination
in the scaling function that could result in double counting when
adding the SuSAM* and the 2p2h cross sections.  The subtracted data
are then used for a fit of a new SuSAM* scaling function to the
$^{12}$C $(e,e')$ data \cite{archive,archive2,Ben08}.  With this new
scaling function we evaluate the total cross section of the SuSAM*+MEC
model and compare with the experimental data.

The scheme of the paper is as follows.
 In Sect. II we present the formalism for the $(e,e')$ cross
section, the SuSAM* response functions and the MEC model in the RMF. 
In sect. III we present the results for the SuSAM*
analysis, the scaling function and the effect of MEC. 
Finally,  in Sect. IV we draw our conclusions.

\section{Formalism}

We are interested in the cross section $\frac{d\sigma}{d\Omega
  d\epsilon'}$ for the interaction of an incident electron with energy
$\epsilon$ that scatters an angle $\theta$ and is detected with final
energy $\epsilon'$. We follow the formalism of Ref. \cite{Ama20}.  
The energy transfer is
$\omega=\epsilon-\epsilon'$ and the four-momentum transfer is
$Q^2=\omega^2-q^2 <0$, where $\nq$ is the (three-) momentum transfer.
Using the plane wave Born approximation with one-photon-exchange the
inclusive cross section is written as
\begin{equation}
\frac{d\sigma}{d\Omega d\epsilon'}
= \sigma_{\rm Mott}
(v_L R_L +  v_T  R_T).
\end{equation}
where $\sigma_{\rm Mott}$ is the Mott cross section, 
 $v_L$ and $v_T$ are  kinematic factors
\begin{eqnarray}
v_L &=& 
\frac{Q^4}{q^4} \\
v_T &=&  
\tan^2\frac{\theta}{2}-\frac{Q^2}{2q^2}.
\end{eqnarray}
The nuclear part of the reaction is contained in 
the longitudinal and transverse response functions, 
 $R_L(q,\omega)$ and $R_T(q,\omega)$, 
respectively. 
\begin{eqnarray}
R_L &=& W^{00}\\
R_T &=& W^{11}+W^{22}
\end{eqnarray}
where $W^{\mu\nu}$ is the hadronic tensor \cite{Ama20}.

\subsection{SuSAM$^*$ response functions}

In this work we compute the cross section as the sum of 
one particle emission (1p1h) plus two-particle emission (2p2h).
The 1p1h part is computed in the super-scaling approach
 with relativistic effective mass (SuSAM*).
This is based in the 
relativistic mean field (RMF) theory of nuclear matter \cite{Ros80,Bar98}.  
In this theory the initial and final nucleons in the 
(1p-1h) excitations are interacting with the nuclear mean field
and acquire an effective mass $m_N^*$.  
The on-shell energy with effective mass is defined as
\begin{equation}
E = \sqrt{p^2+(m_N^*)^2}
\end{equation}
Note that this is not the true energy of the nucleon in the RMF, but
in the particular case of 1p-1h excitations, the responses only depend
on the differences between initial and final energies, and therefore
only the on-shell energy appears (the extra vector energy will be discussed in 
Sect. IIC).
In the mean field the initial nucleon
has momentum $p$ below the  Fermi momentum, $p < k_F$.
and  on-shell energy $E=\sqrt{\np^2+m_N^*{}^2}$
The final nucleon has momentum
$\np'=\np+\nq$, and the final on-shell energy is
$E'=\sqrt{\np'{}^2+m_N^*{}^2}$.  Pauli blocking implies $p' > k_F$.

Similarly to the RFG, the nuclear response functions in the RMF can be written 
as the product of an averaged 
 single nucleon response times the scaling function \cite{Mar17,Ama18}
\begin{equation}
R_K(q,\omega)  =   r_K(q,\omega) f^*(\psi^*),  
\kern 1cm K=L,T \label{factorization}  
\end{equation}
where the single-nucleon responses $r_L$ and $r_T$ are
given below.
The scaling function for nuclear matter is  
\begin{equation} 
f^*(\psi^*)= \frac34(1-\psi^{*2})\theta(1-\psi^{*2})
\end{equation}
where the scaling variable $\psi^*$ is defined as
follows
\begin{equation}
\psi^* = \sqrt{\frac{\epsilon_0-1}{\epsilon_F-1}} {\rm sgn} (\lambda-\tau).
\end{equation}
Where $\epsilon_0$ is the minimum energy allowed for the initial nucleon 
absorbing the energy and momentum transfer $(\omega,q)$,
in units of $m_N^*$, given by
\begin{equation}
\epsilon_0={\rm Max}
\left\{ 
       \kappa\sqrt{1+\frac{1}{\tau}}-\lambda, \epsilon_F-2\lambda
\right\},
\end{equation}
In the above formulas we have used the following 
dimensionless variables 
\begin{eqnarray}
\lambda  &=& \omega/2m_N^*, \\
\kappa  & = & q/2m_N^*,\\
\tau & = & \kappa^2-\lambda^2, \\
\eta_F & = &  k_F/m_N^*,\\
\xi_F & = & \sqrt{1+\eta_F^2}-1,\\
\epsilon_F &=& \sqrt{1+\eta_F^2}, 
\end{eqnarray}
Note that all these variables are modified with respect to the RFG, by
including the effective mass instead of the free nucleon mass.

The single-nucleon response functions are defined as
\begin{equation}
r_K = \frac{\xi_F}{m^*_N \eta_F^3 \kappa} (Z U^p_K+NU^n_K)
\label{single} 
\end{equation}
Where $Z$ ($N$) is the number of protons (neutrons).
The 
functions $U_L, U_T$ 
are given by
\begin{eqnarray}
U_L &=& \frac{\kappa^2}{\tau}
\left[ (G^*_E)^2 + \frac{(G_E^*)^2 + \tau (G_M^*)^2}{1+\tau}\Delta \right]
\\
U_T &=& 2\tau  (G_M^*)^2 + \frac{(G_E^*)^2 + \tau (G_M^*) ^2}{1+\tau}\Delta
\end{eqnarray}
where 
\begin{equation}
\Delta= \frac{\tau}{\kappa^2}\xi_F(1-\psi^*{}^2)
\left[ \kappa\sqrt{1+\frac{1}{\tau}}+\frac{\xi_F}{3}(1-\psi^*{}^2)\right].
\end{equation}
Finally,
the electric and magnetic form factors in the RMF 
are \cite{Ama15, Bar98}:
\begin{eqnarray}
G_E^*  &=&  F_1-\tau \frac{m^*_N}{m_N} F_2 \label{GE} \\
G_M^*  &=& F_1+\frac{m_N^*}{m_N} F_2.  \label{GM}
\end{eqnarray}
where 
 $F_1$ and $F_2$, 
are the Dirac and Pauli form
factors from the 
electromagnetic current
operator \cite{For83}
\begin{equation}
J^\mu_{s's}=
\overline{u}_{s'}(\np')
\left[ 
F_1\gamma^\mu 
+F_2i\sigma^{\mu\nu}\frac{Q_\nu}{2m_N}
\right]u_{s}(\np)
\label{vector}
\end{equation}
For the $F_i$ form factors of the nucleon, we use the Galster
parameterization \cite{Gal71}.  Note that in Eq. (\ref{GE}) the
variable $\tau=|Q^2|/(4m_N^{*2})$ is modified also in the medium the
effective mass instead of the free nucleon mass.

We have so far described the response functions of the RMF theory of
nuclear matter. In the SuSAM* approach we assume that the
factorization, Eq. (\ref{factorization}), is approximately valid for finite nuclei, but the scaling function is modified by a 
phenomenological function extracted from experimental data in the next section, and that is parametrized in the following way
\begin{equation} \label{central}
f^*(\psi^*) = 
 a_3e^{-(\psi^*-a_1)^2/(2a_2^2)}+ b_3e^{-(\psi^*-b_1)^2/(2b_2^2)}.
\end{equation}
On the contrary to the RMF, the SuSAM* scaling function is not zero
outside the interval $-1< \psi^*< 1$, providing extra contributions to
the cross section for low and large values of the scaling variable
not present in nuclear matter models.

\begin{figure*}
\includegraphics[width=11cm, bb=100 450 520 700]{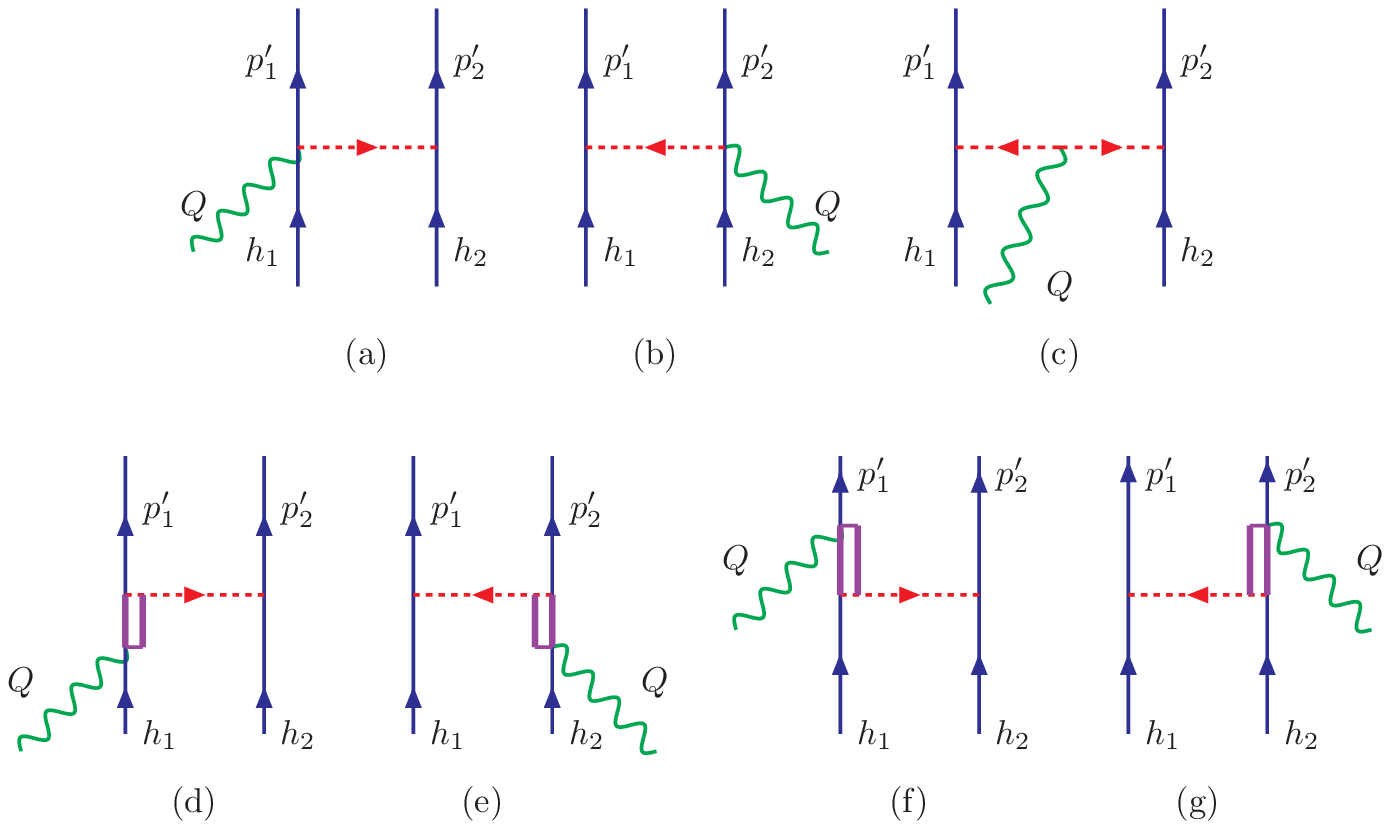}
\caption{ 
Feynman diagrams for the electromagnetic MEC used in this work.
}
\label{diagmec}
\end{figure*}

\subsection{2p2h responses and MEC}

In this work we apply a fully relativistic model of meson exchange
currents (MEC) to compute the electromagnetic response functions in the
two-nucleon emission channel (2p-2h). The model was developed for the
RFG in Ref. \cite{Sim16}.

\begin{widetext}

The 2p-2h hadronic tensor is computed by integrating over all
the 2p-2h excitations 
 \begin{eqnarray}
W^{\mu\nu}_{\rm 2p2h}
&=& 
\frac{V}{(2\pi)^9}\int
d^3p'_1
d^3p'_2
d^3h_1
d^3h_2
\frac{m_N^4}{E_1E_2E'_1E'_2}
w^{\mu\nu}(\np'_1,\np'_2,\nh_1,\nh_2)\;
\delta(E'_1+E'_2-E_1-E_2-\omega)
\nonumber\\
&&
\times 
\Theta(p'_1,h_1)\Theta(p'_2,h_2)
\delta(\np'_1+\np'_2-\nq-\nh_1-\nh_2) \, ,
\label{amaro-hadronic12}
\\
&&
\kern -1cm = \frac{V}{(2\pi)^9}\int
d^3p'_1
d^3h_1
d^3h_2
\frac{m_N^4}{E_1E_2E'_1E'_2}
\Theta(p'_1,h_1)\Theta(p'_2,h_2)
w^{\mu\nu}(\np'_1,\np'_2,\nh_1,\nh_2)\;
\delta(E'_1+E'_2-E_1-E_2-\omega) 
\label{amaro-hadronic}
\end{eqnarray}
where 
$\nh_1$ and $\nh_2$ are the momenta of the holes, with $h_i < k_F$, 
while $\np'_1$ and $\np'_2$ are the momenta of the final particles
with $p'_i> k_F$  (Pauli blocking). These two conditions are enforced by the  
functions $\Theta$ appearing inside the integral, defined as 
the product of step-functions
$\Theta(p',h) \equiv
\theta(p'-k_F)
\theta(k_F-h)$.

\end{widetext}

In Eq. (\ref{amaro-hadronic}) we have integrated
 over $\np'_2$ using the delta-function of momentum conservation
with $\bf p'_2= h_1+h_2+q-p'_1$.  

The hadronic tensor for a single 2p2h excitation,
 $w^{\mu\nu}(\np'_1,\np'_2,\nh_1,\nh_2)$, 
is defined as
\begin{eqnarray}
&& w^{\mu\nu}(\np'_1,\np'_2,\nh_1,\nh_2) =
\nonumber \\
&&
\frac{1}{4}
\sum_{s_1s_2s'_1s'_2}
\sum_{t_1t_2t'_1t'_2}
j^{\mu}(1',2',1,2)^*_A
j^{\nu}(1',2',1,2)_A
\label{amaro-elementary}
\end{eqnarray}
where sums are performed over over spin and isospin third components.
Here, the  2p-2h electromagnetic current matrix element is defined as
\begin{equation} 
j^\mu(1',2',1,2)
\equiv j^\mu(\np'_1s'_1t'_1,\np'_2s'_2t'_2,\nh_1s_1t_1,\nh_2s_2t_2)
\end{equation}
and the sub-index $A$ means 
direct minus exchange matrix element
\begin{equation} \label{amaro-anti}
j^{\mu}(1',2',1,2)_A
\equiv j^{\mu}(1',2',1,2)-
j^{\mu}(1',2',2,1) \,.
\end{equation}
The above integral can be reduced to a 7 dimension integral using the energy conservation function and axial symmetry. Details are given in ref.
\cite{Sim14}.

The model of relativistic meson-exchange currents correspond to the
sum of the Feynman diagrams of Fig. 1 \cite{DeP03,Sim16}.  Diagrams
(a,b) represent the seagull (or contact) current, diagram (c) is the
pion in flight current, and diagrams (d,e) are the $\Delta$ forward
current and (f,g) the $\Delta$ backward current.  The total
$\Delta(1232)$ excitation current is the sum of forward plus backward
diagrams. Specifically the MEC matrix element is written as
\begin{equation}
j^\mu(1',2,1,2) = 
j_{\rm sea}^\mu +j_{\rm \pi}^\mu + j_{\rm \Delta}^\mu
\end{equation}
The three two-body currents: i) seagull, 
$j_{\rm sea}^\mu$, ii) pion-in-flight, $j_{\rm \pi}^\mu$,  and iii) 
$\Delta$ isobar, $j_{\rm \Delta}^\mu$, are defined next.

\begin{widetext}
The seagull current matrix element (diagrams (a) and (b) in Fig. 1) is 
 \begin{equation}
 j^\mu_{\rm sea}=
\frac{f^2}{m^2_\pi}
  \left[I_V^{3}\right]_{1'2',12}
V_{\pi NN}^{s'_1s_1}(\np'_1,\nh_1) 
 \bar{u}_{s^\prime_2}(\np^\prime_2)
 F^V_1(Q^2)F_{\pi NN}(k_1^2)
\gamma_5 \gamma^\mu
  u_{s_2}(\nh_2)
   +
   (1\leftrightarrow2) \,
\label{seacur}
\end{equation}
\end{widetext}
where the $\pi NN$ 
 coupling constant is $f=1$,
we use the following two-body isospin operator 
\begin{eqnarray}
I_V^3  & =&  i \left[\tauvec(1) \times\tauvec(2)\right]_z .
\end{eqnarray}
The $\pi NN$ vertex and the pion propagator appear in the 
 spin-dependent function
\begin{equation}
V_{\pi NN}^{s'_1s_1}(\np'_1,\nh_1) \equiv 
F_{\pi NN}(k_1^2)
\frac{\bar{u}_{s^\prime_1}(\np^\prime_1)\,\gamma_5
 \kbar_{1} \, u_{s_1}(\nh_1)}{k^2_{1}-m^2_\pi},
\end{equation}
where $k_i^\mu=p'_i{}^\mu-h_i{}^\mu$, $i=1,2$, 
 is the momentum transfer to the $i$th particle, $m_\pi$ is the pion mass, and 
$F_{\pi NN}$ is the strong $\pi NN$ form factor
\cite{Som78,Alb84}
\begin{equation}
F_{\pi NN}(k_1^2)= \frac{\Lambda^2-m_\pi^2}{\Lambda^2-k_1^2}
\end{equation}
where we use $\Lambda=1300$ MeV.
Finally  $F_1^V=F_{1p}-F_{1n}$ 
is  the isovector electromagnetic form factor of the nucleon.

\begin{widetext}
The pion-in-flight or pionic current matrix element follows from
diagram (c) in Fig. 1, given as
 \begin{equation}
 j^\mu_{\pi}=
 \left[I_V^{3}\right]_{1'2',12}
 \frac{f^2}{m^2_\pi}
 F^V_1(Q^2)
V_{\pi NN}^{s'_1s_1}(\np'_1,\nh_1) 
V_{\pi NN}^{s'_2s_2}(\np'_2,\nh_2) 
\left(k^\mu_{1}-k^\mu_{2}\right).
\label{picur}
\end{equation}
Finally, the $\Delta$ current is the sum of forward (diagrams d,e) 
and backward (f,g) contributions
 \begin{eqnarray*}
j^\mu_{\Delta}
&=&
\frac{f^* f}{m^2_\pi}\,
V_{\pi NN}^{s'_2s_2}(\np'_2,\nh_2) 
\bar{u}_{s^\prime_1}(\np^\prime_1)
F_{\pi N\Delta}(k_2)
\left\{ 
\left[U_{\rm F}^{3}\right]_{1'2',12}
 k^\alpha_{2}
G_{\alpha\beta}(h_1+Q)
\Gamma^{\beta\mu}(h_1,Q)
\right.
\nonumber\\
&& +
\left.
\left[U_{\rm B}^{3}\right]_{1'2',12}\; 
k^\beta_{2}
\hat{\Gamma}^{\mu\alpha}(p^\prime_1,Q)
G_{\alpha\beta}(p^\prime_1-Q)
\right\}
u_{s_1}(\nh_1)
+(1\leftrightarrow2)
\label{deltacur}.
\end{eqnarray*}
\end{widetext}

We use the $\pi N\Delta$ coupling
constant $f^*=2.13$. 
The forward and backward isospin transition operators
are defined by
\begin{eqnarray}
U_{\rm F}^3&=&\sqrt{\frac32}
\sum_i\left(T_i T_3^\dagger\right)\otimes
\tau_i
\label{forward}
\\
U_{\rm B}^3&=&
\sqrt{\frac32}
\sum_i\left(T_{3}\, T^\dagger_i\right)\otimes
\tau_i,
\label{backward}
\end{eqnarray}
where $\vec{T}$ is the transition
operator from isospin $\frac32$ to $\frac12$.

We use the $\pi N \Delta$ strong form factor of Ref. \cite{Dek94}, given by
\begin{equation}
F_{\pi N\Delta}(k_2^2)=
 \frac{\Lambda^2_\Delta}{\Lambda_\Delta^2-k_2^2}
\end{equation}
were $\Lambda_\Delta=1150$ MeV. 

For the electromagnetic
$N\rightarrow\Delta$ transition tensor
in the forward current, $\Gamma^{\beta\mu}(P, Q)$ we use
\begin{equation}
\Gamma^{\beta\mu}(P,Q)=
\frac{C^V_3}{m_N}
\left(g^{\beta\mu}\Qbar-Q^\beta\gamma^\mu\right)\gamma_5.
\end{equation}
We have kept only the $C_3^V$  form factor and neglected the 
smaller contributions of the higher-order terms
to the interaction of Ref. \cite{Hernandez:2007qq}.
In the backward current, the vertex tensor is
\begin{equation}
\hat{\Gamma}^{\mu\alpha}(P^\prime, Q)=\gamma^0
\left[\Gamma^{\alpha\mu}(P^\prime,-Q)\right]^{\dagger}
\gamma^0 \, .
\end{equation}

Finally, for the $\Delta$-propagator we use
\begin{equation}\label{delta_prop}
 G_{\alpha\beta}(P)= \frac{{\cal P}_{\alpha\beta}(P)}{P^2-
 M^2_\Delta+i M_\Delta \Gamma_\Delta(P)+
 \frac{\Gamma_{\Delta}(P)^2}{4}} \, .
\end{equation}
where ${\cal P}_{\alpha\beta}(P)$ is 
the spin-$\frac32$ projector  
\begin{eqnarray}
{\cal P}_{\alpha\beta}(P)&=&-(\Pbar+M_\Delta)
\left[g_{\alpha\beta}-\frac13\gamma_\alpha\gamma_\beta-
\frac23\frac{P_\alpha P_\beta}{M^2_\Delta}\right.
\nonumber\\
&+&\left.
\frac13\frac{P_\alpha\gamma_\beta-
P_\beta\gamma_\alpha}{M_\Delta}\right].
\end{eqnarray}
For the $\Delta$-width, $\Gamma_\Delta(P)$ we use the prescription of Dekker
\cite{Dek94}.

\subsection{Inclusion of MEC in the RMF model}

In past works \cite{Sim16,Meg16b,Sim17}, the 2p2h responses have been
computed with the formalism of the previous subsection in the RFG model,
including an energy shift $\omega \rightarrow \omega - E_S$ to take
into account the separation energy of two nucleons $E_S \simeq 40$ MeV
of finite nuclei that cannot be described in the Fermi gas. This shift
is not applied in the electromagnetic form factors of the currents.

In this work we modify the above MEC model for consistency with the
relativistic mean field (RMF) of nuclear matter in which the SuSAM$^*$
formalism is based. In the RMF the nucleon interacts with the
self-consistent mean field in the Hartree approximation (Walecka
model), and acquires scalar and vector potential energies. The scalar
energy gives rise to the nucleon effective mass
\begin{equation}
m_N^*= m_N - g_s\phi_0 = M^* m_N
\end{equation}
where $g_s\phi_0$ is the scalar potential energy that depends
explicitly on the scalar field $\phi_0$ of the RMF \cite{Ser86}.  On
the contrary the vector field, $V_0$, of the RMF produces a repulsive
potential, $E_v=g_vV_0$, or vector energy, that is added to the
on-shell energy to obtain the true nucleon energy
\begin{equation}
E_{true}= E + E_v
\end{equation}
where $E = \sqrt{p^2+(m_N^*)^2}$ is the on-shell energy with effective
mass $m_N^*$.

The SuSAM* model is inspired by the 1p-1h quasielastic responses of
the RFG, where only differences of energies between initial and final
nucleons appear. Therefore the vector energy $E_v$ cancels and does
not appear in the 1p-1h cross section, and the resulting responses are
computed as in the RFG with the change $m_N\rightarrow m_N^*$, except
for the electromagnetic current operator \cite{Mar17,Ama18}. 
 This change has already
been specified in the equations of the previous section.

The same cancellation happens in the case of the 2p-2h seagull and
pionic current matrix element because the vector energy cancels in the
time components of the vectors $k_1=p'_1-h_1$ and $k_2=p'_2-h_2$.
However, in the case of the $\Delta$ current, the $\Delta$ propagator
must be computed with the true nucleon energy, including the vector
energy. Thus in the forward propagator, $G_{\alpha\beta}(h+Q)$, the
hole energy must be the true nucleon energy $h_0=E_h+E_v=
(h^2+m_N^{*2})^{1/2}+E_v$.  The inclusion of the vector energy affects
to the position of the pole in the forward $\Delta$ diagrams, giving
rise to the $\Delta$ peak. This allows to determine the value of the
vector energy from the experimental data.  The same modification must
be also applied to the backward propagator $G_{\alpha\beta}(p'-Q)$,
and to the electromagnetic vertices $\Gamma^{\mu\nu}(h,Q)$ and
$\hat{\Gamma}^{\mu\nu}(p',Q)$, although in our case these electromagnetic
vertices only depend on $Q$.

To finish the implementation of MEC in the RMF we modify the nucleon
spinors by using the relativistic effective mass $m^*_N$ instead of
$m_N$ in all places except in the form factor $C_3^V/m_N$.  All the
remaining energies in the hadronic tensor (\ref{amaro-hadronic}) are
modified accordingly with the on-shell energy of nucleons with effective
mass $m_N^*$, and the vector energy cancels.

With this procedure we already have at our disposal a consistent model
of quasielastic (SuSAM*) plus 2p-2h response functions with
relativistic effective mass and vector energy. Note that the use of
the effective mass accounts for the nucleon separation energy in both
the quasielastic and 2p2h channels.

In the next section we will compare both 2p2h models with and
without effective mass, and will use those model to subtract the
2p2h channel to the experimental data of electron scattering.

\begin{figure}
\includegraphics[width=7cm, bb=120 140 430 780]{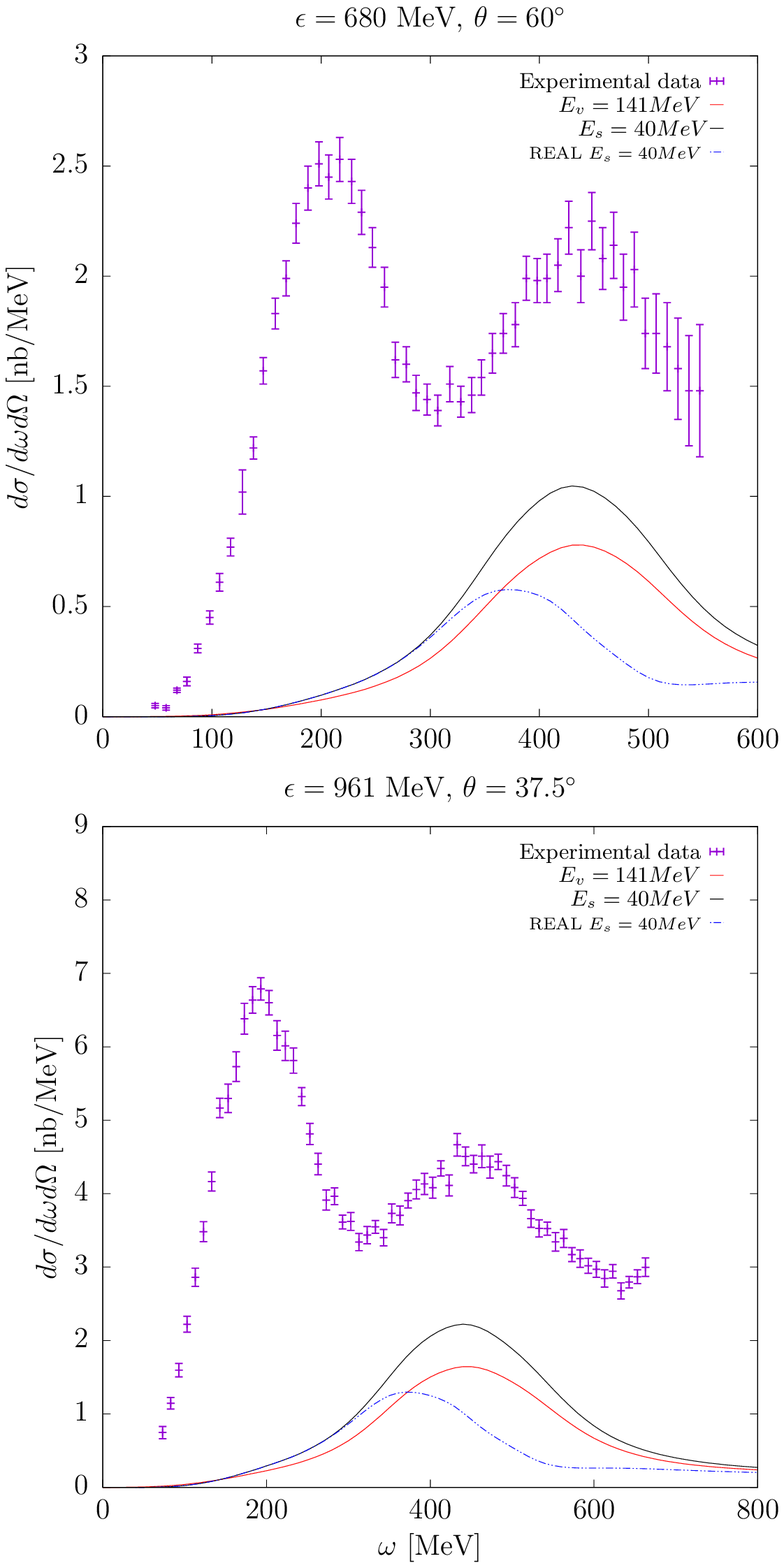}
\caption{ 2p2h contribution to the cross section of
  $^{12}$C compared to $(e,e')$ data for two experimental
  kinematics. Black lines are the RFG results with separation energy
  $E_s=40$ MeV, and blue lines are the same model with only the real
  part of the $\Delta$ propagator. In red are the RMF results with
  $M^*=0.8$ and vector energy $E_v=141$ MeV.  The experimental data
  from Ref. \cite{archive,archive2,Ben08}
 }
\label{revalida2}
\end{figure}

\section{Results}


\subsection{Obtaining the vector energy}

In Fig. 2 we compare the 2p-2h inclusive cross section of the RFG
model (with $M^*=1$ and separation energy $E_s=40$ MeV) to the RMF
model (with $M^*=0.8$ and $E_v=141$ MeV) for electron scattering from
$^{12}$C and two different kinematics. Also shown is the calculation
corresponding to the model of ref. \cite{Meg16b}, where the RFG 2p2h
responses were computed using the real part of the denominator of the $\Delta$
propagator, producing smaller cross section peaking at the dip region
between the quasielastic and delta peak. 

We highlight that in this work we instead use the total delta
propagator (real plus imaginary parts) that produces a peak centered
in the delta resonance region. This peak describes $\Delta$ excitation
decaying inside the nucleus with two nucleon emission instead of pion
emission.  This decay channel of the $\Delta$ is superposed to the
pion emission peak because the same delta propagator contributes to
both processes.  The 
differences in the strength of the 2p2h in the two models of fig. 2 
with full $\Delta$ propagator (black and red lines) is
a result of the relativistic mean field included in the red lines.

The effect of the effective mass is a reduction of the 2p-2h peak and
a shift in $\omega$. This shift is not shown in the figure because it
cancels out by the vector energy, $E_v=141$ MeV, in the $\Delta$
propagator.  This value of the vector energy has been fixed such that
the $\Delta$ peak in the 2p2h cross section is at the same position as
the $\Delta$ resonance of the experimental data.

In fact the maximum of the forward $\Delta$ propagator occurs
approximately for $(H+Q)^2-M_\Delta^2=0$. For a nucleon at rest,
$\nh=0$, in the RFG, this implies that
\begin{equation} \label{pico1}
\omega-E_S = \sqrt{M_\Delta^2+q^2}-m_N
\end{equation}
where $E_S \simeq 40$ MeV represents the separation energy of two-nucleon emission that has to be subtracted to the energy transfer.
On the other hand, in the RMF model, the condition is
\begin{equation} \label{pico2}
\omega = \sqrt{M_\Delta^2+q^2}-m^*_N-E_v
\end{equation}
where in this case the separation energy is not needed because it is
implicitly included in the scalar potential that gives rise to the
relativistic effective mass $m_N^*= m_N - g_s\phi_0$.
Comparing Eqs. (\ref{pico1}) and (\ref{pico2}) we obtain 
\begin{equation}
m_N-E_S = m_N^*+E_v
\end{equation}
from where $E_v = m_N-m_N^*-E_S \simeq 148$ MeV. This estimated value
is in agreement with the fitted value $E_v=141$ MeV used in Fig. 2.

The sum of scalar plus vector energy gives
the total potential energy of the nucleon
\begin{equation}
U_0 \equiv -g_s\phi_0 +g_vV_0 = (m^*_N-m_N) +E_v
\end{equation}
In our case, $M^*=0.8$, 
using the fitted value, $E_v=141$ MeV, 
this gives $U_0 \simeq -47$ MeV 
for the depth of the nucleon potential energy in $^{12}$C from our 
2p2h model.

This provides a procedure to obtain the vector energy from electron
scattering data as a function of the effective mass. We can compare
with the values obtained by the model of Serot-Walecka in
ref. \cite{Ser86}, where $E_v = 330$ MeV for effective mass $M^*=0.6$
in nuclear matter. The corresponding scalar potential is
$g_s\phi_0=376$ MeV. And the depth of the total potential is $U_0
\simeq -46$ MeV, in good agreement with our findings.

A similar reduction effect of the 2p2h peak due to the relativistic
mean field is obtained for the other kinematics. This reduction
amounts to $\sim 25\%$ of the RFG model.

\begin{figure*}
\includegraphics[width=15cm, bb=31 273 534 780]{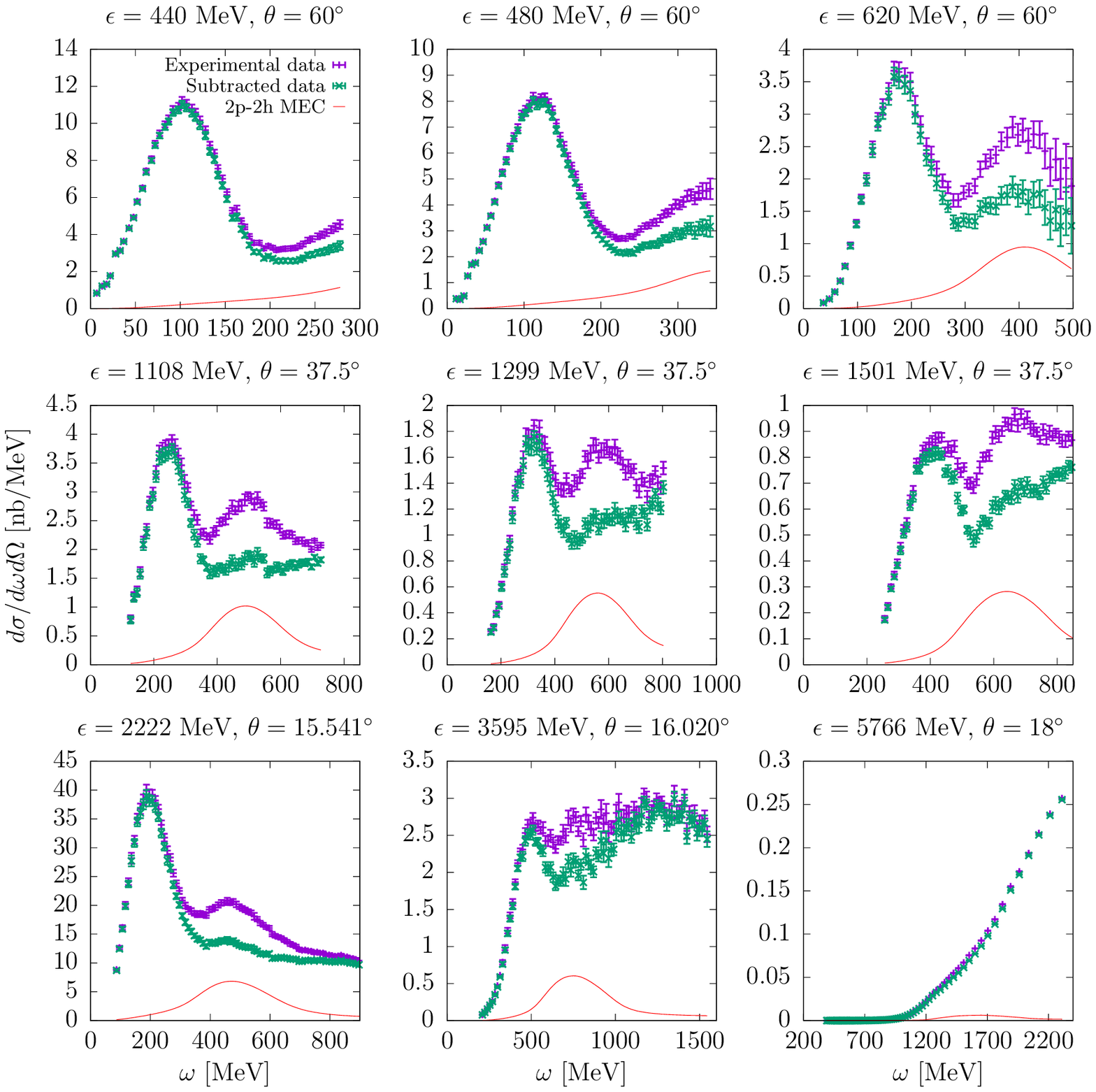}
\caption{ Experimental and subtracted electron scattering data for
  selected kinematics.  The solid lines are the 2p2h cross sections
  values that are being subtracted, computed with the RMF model with
  $M^*=0.8$.  Data from \cite{archive,archive2} }
\label{revalida1}
\end{figure*}

\subsection{Subtraction of 2p2h cross section from data}

Once we have obtained the phenomenological vector energy for our RMF
model of 2p2h response, the next step is to subtract the 2p2h contribution
from the experimental electron scattering data. The reason is to
obtain a better description of the quasielastic scaling function in
the SuSAM* model without the 2p2h MEC contamination. Therefore the
subtracted data should be a more reliable representation of the 1p1h
excitations, and therefore they are more appropriate to be used as a
starting point to perform a scaling analysis. 
Therefore the subtracted data are defined as
\begin{equation}
\left(\frac{d\sigma}{d\Omega' d\epsilon'}\right)_{sub}\equiv
\left(\frac{d\sigma}{d\Omega' d\epsilon'}\right)_{exp}-
\left(\frac{d\sigma}{d\Omega' d\epsilon'}\right)_{2p2h}
\end{equation}
One of the aims of this work is to perform a scaling analysis of the
subtracted data, similar to the ones performed in refs. \cite{Ama15,Ama18}.

First, in Fig. 3 we show the result of the subtraction for some
kinematics of $^{12}$C.  The 2p2h curve that is being subtracted from
data contributes mainly in the region to the right of the quasielastic
peak. Therefore, the subtraction does not modify largely the data
around the quasielastic region. The larger effect occurs in the
resonance $\Delta$ peak, that is dominated by pion emission. Note that
the 2p2h cross section that is being subtracted is not a contribution to
pion emission because the final states are two nucleons in the
continuum. As we will see below, the contribution of the data in the
inelastic region will be irrelevant to the scaling analysis and do not
influence the extraction of the quasielastic scaling function.

In Fig. 3 we use the
RMF model with $M^*=0.8$. We have also made the subtraction with the
RFG model (not shown in the figure), where the reduction of the data
is larger.

We have performed this subtraction for all the available data of
$^{12}$C. In total there are 2969 entries in the data base. We will
use the resulting subtracted data to perform the scaling analysis in
the next subsection.

\begin{figure}
\includegraphics[width=7cm, bb=145 440 393 775]{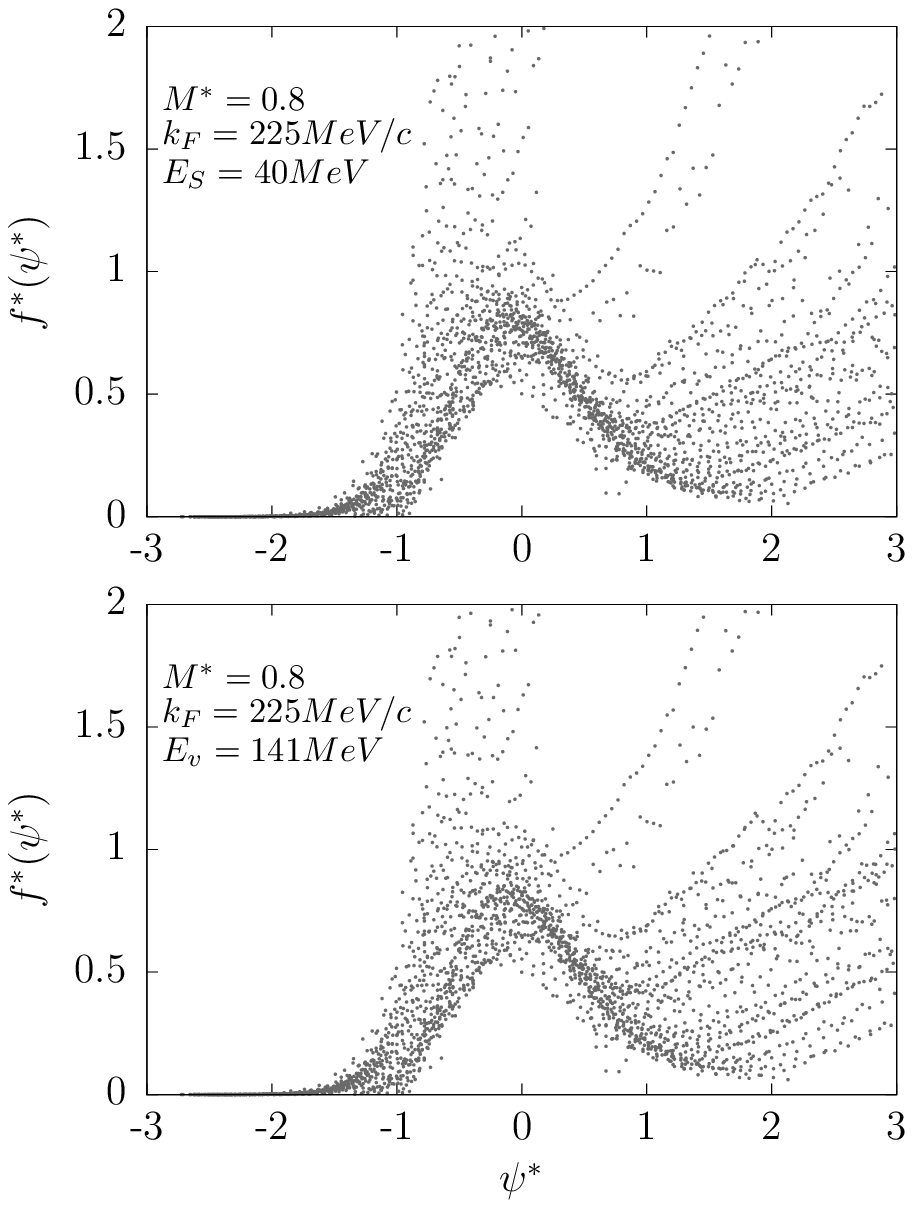}
\caption{ Scaling plot of the subtracted cross section data scaled
  with the single nucleon function displayed as a function of the
  scaling variable $\psi^*$ using effective mass $M^*=0.8$ and Fermi
  momentum $k_F=225$ MeV/c.  In the top panel the 2p2h has been
  computed in the RFG with separation energy $E_s=40$ MeV. In the
  bottom panel the 2p2h are computed in the RMF model with the same
  value of $M^*=0.8$.
}
\label{revalida3}
\end{figure}

\subsection{Scaling analysis of subtracted data}

We have developed in the past several SuSAM* models obtained by
different methods to perform the scaling analysis of $(e,e')$ cross
section data without the subtraction of the 2p2h.  We carried out
such analyses in refs. \cite{Ama15,Ama17,Mar17,Ama18}.  In the SuSAM*
models the response functions are computed by using Eq. (7),
replacing the RFG scaling function, Eq (8), by a phenomenological function
$f^*(\psi^*)$, that is parametrized by Eq. (24).  
The SuSAM* procedure has proven to be quite robust
in the sense that
different methods produce similar results for the SuSAM* parameters,
verifying self-consistency and superscaling, i.e. that the same
scaling function is valid for all the nuclei studied.

In this section from the subtracted experimental data we obtain a new
phenomenological scaling function $f^*(\psi^*)$ without the
contamination of 2p2h states.  One of the goals of this work is to
quantify the change of the scaling function due to this subtraction.

To obtain the new scaling function, we first compute
for every subtracted datum 
 the experimental value of the scaling function
$f^*$ by dividing the subtracted cross section by the
single nucleon contribution
\begin{equation}
f^* =
\frac{\left(\frac{d\sigma}{d\Omega'd\epsilon'}\right)_{\rm sub}}{
  \sigma_{\rm Mott}\left( v_L r_L + v_T r_T \right)}
\end{equation}
 We also compute the corresponding value of the scaling variable
 $\psi^*$ for that datum.  In Fig. 4 we plot $f^*$ versus $\psi^*$ for
 all the data of $^{12}$C.  The values of $M^*=0.8$ and Fermi momentum
 $k_F=225$ MeV/c have been taken from the previous analyses of
 refs. \cite{Ama15,Ama17,Mar17,Ama18}, where it was shown that these
 values provide the best scaling of data.  

In Fig. 4 we show the subtracted results using the two models of 2p2h
discussed above.  In the top panel of Fig. 4 the MEC have been
computed in the RFG with $M^*=1$ and a separation energy. In the
bottom panel, the MEC have been computed in the RMF with the same
value of the effective mass $M^*=0.8$ used to compute the scaling
function.  In the first case there is an inconsistency because we are
using two different values for the nucleon mass: $m_N^*$ is being used
to compute the scaling function and the scaling variable, while $m_N$
is being used to compute the MEC 2p2h subtraction.  The second case is
consistent because we are using always the same value for the nucleon
mass, $m_N^*$. However in both cases the resulting scaling plot is
very similar, because both MEC models differ in $\simeq 25$ \%.

The most striking thing about the graphs in Fig. 4 is that many points
accumulate forming a narrow band or point cloud. This band can be
extracted if we eliminate the most scattered data from the plot.  To
discard the data we apply a density criterion by computing the
density of the points in the plot and keeping only those points
surrounded by more than 25 points within a circle of radius $r=0.1$.
This is the same criterion used in our previous works
 \cite{Ama15,Ama17,Mar17,Ama18}.

\begin{figure}
\includegraphics[width=7cm, bb=145 440 393 775]{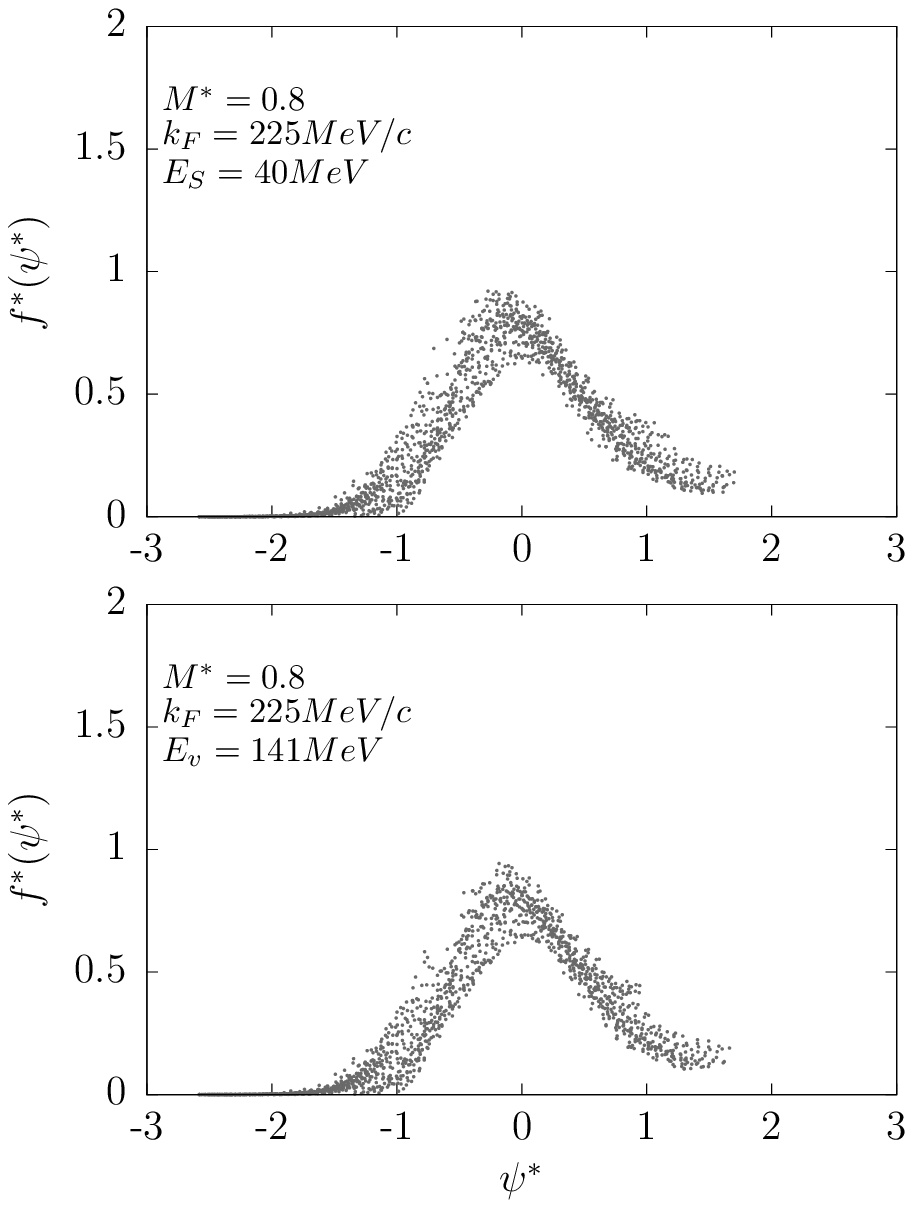}
\caption{Scaling analysis of the subtracted cross section data after 
discarding the more scattered data by a density criterion.
The selected data are shown for the two models of the 2p2h cross section
of Fig. 4.}  
\label{revalida4}
\end{figure}

In Fig. 5 we show the selected data resulting from the application of
the density criterion. All the inelastic data points go away and 
only the points around the QE region survive as a thick band.  In past
works we obtained the QE bands without subtraction of the 2p2h cross
section. In the present work the 2p2h contribution is not present in
the data points.  Besides, we observe that both bands are very similar
in both MEC models.  In the top case (RFG) a total of 1546 data points
from the total 2969 points survive.  In the bottom case (RMF) the band
contains 1453 data points. However, both bands are almost identical.
Note that the selected points accumulate around the values of the
scaling function of the RMF (for nuclear matter) $f_{RMF}=
3(1-\psi^{*2})\theta(1-\psi^{*2})/4$. However the cloud of data points
extend outside the interval $-1 < \psi^* <1$, where the RMF scaling
function is zero. Specifically, most of the data points are in the
range $-1.5 < \psi^* < 1.5$.

Note that the density criterion is one choice to approximate the
region where more data collapse, and where a band is
clearly visible with defined edges. This choice provides 
an estimate of the degree
of scaling violation, from the width of the resulting band,
because it is clear that the data do not scale
exactly, but only approximately.

\begin{table*}

\begin{tabular}{ccrrrrrr}\hline
 &  & $a_1$ & $a_2$ & $a_3$ & $b_1$ & $b_2$ & $b_3$ \\ \hline

&central 
&-0.0465
& 0.469
& 0.633
& 0.707
& 1.073
& 0.202
\nonumber\\
Old band
&min
&-0.0270
& 0.442
& 0.598
& 0.967
& 0.705
& 0.149
\nonumber\\

&max
& -0.0779
& 0.561
& 0.760
& 0.965
& 1.279
& 0.200
\\ 
\\
&central 
& -0.0971       
& 0.422         
& 0.477         
& 0.299         
& 0.855        
& 0.330      
\nonumber\\
New band
&min
& -0.0419      
& 0.437        
& 0.575       
& 0.759      
& 0.625       
& 0.152 
\nonumber\\

&max
& -0.1594        
& 0.585        
& 0.759        
& 0.863         
& 0.965        
& 0.230    
\\ \hline
\end{tabular}

\caption{Parameters of the central value of 
 the phenomenological scaling function, $f^*(\psi^*)$, 
 and those of the lower and upper boundaries (min and
 max, respectively) of the bands.
 }
\label{bandas}
\end{table*}

\begin{figure}
\includegraphics[width=8cm, height=8cm, angle=-90]{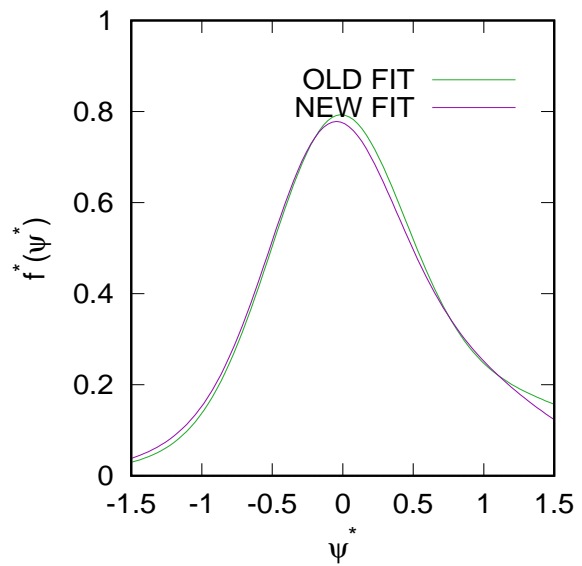}
\caption{
Comparison between the phenomenological scaling function obtained in this work
with 2p2h subtraction (new fit) and the scaling function obtained in 
ref. \cite{Ama17} without subtraction (old fit). }
\label{revalida5}
\end{figure}

The phenomenological scaling function in the subtracted SuSAM$^*$ model 
is defined by a fit to the selected data points, with a function  
that we parameterize as a combination of two Gaussian functions, Eq. (24).
This provides the central value of the band.  This scaling function
gives the best approximation to the quasi-elastic data in the SuSAM*
model. The resulting new scaling function is shown in Fig. 6, where it
is compared with the old scaling function obtained without
subtracting the 2p-2h \cite{Ama17}. The parameters of the new and old
fits are given in Table \ref{bandas}.

Concerning the theoretical error coming from scaling violation in the
quasielastic data of Fig. 5, we estimate it by fitting the maximum and
minimum values in the point cloud. This is done by choosing an enough
small bin size in $\psi^*$ in order to determine a subset of points defining
the experimental borders of the band.  We divide the interval of the
variable $\psi^*$ into sub-intervals of width $\epsilon$ (bins).
Within each bin of $\psi^*$ we calculate the maximum and minimum of
the scaling function of all the points within the bin. These maximums
and minimums define the points of the upper and lower edges.  These
edges are fitted separately as sums of two gaussians similarly to Eq.
(24).  The resulting theoretical band is shown in Fig. 7,
where it is compared to the band fitted without subtraction of 2p2h
\cite{Ama17}.  The central scaling function previously fitted provides
the best approximation to the selected data points, and therefore to
the quasielastic cross section without 2p2h, within a theoretical
error given by the band.

Note that the bands of Fig. 7, fitted with and without 2p2h
contribution, are very similar. The small differences are due to the
slight change of some data after subtraction, but the quasielastic
region defined by them is unaltered by MEC.  This is because the data
that are most affected by MEC are those that are later removed by
the selection process. 
These results confirm that the SuSAM* approach to select the QE data
 is  consistent with or without the subtraction.
 What we have achieved with subtraction is a better definition 
of the tail to the right of the scaling function, 
which extends above $\psi^* = 1.5$. 

\begin{figure}
\includegraphics[width=6cm, angle=-90]{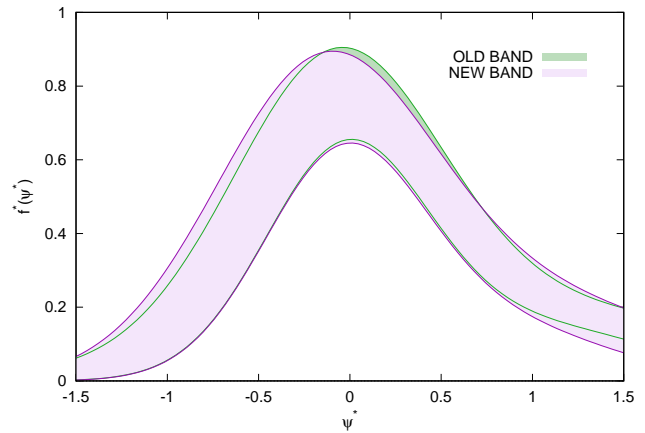}
\caption{
The quasielastic band of the scaling function obtained  in the present work with subtraction of 2p2h cross section, 
compared to the old band fitted in ref. \cite{Ama17} without subtraction.}
\label{revalida5}
\end{figure}

\subsection{Cross section results}

\begin{figure*}
\includegraphics[width=18cm, bb=35 270 535 770]{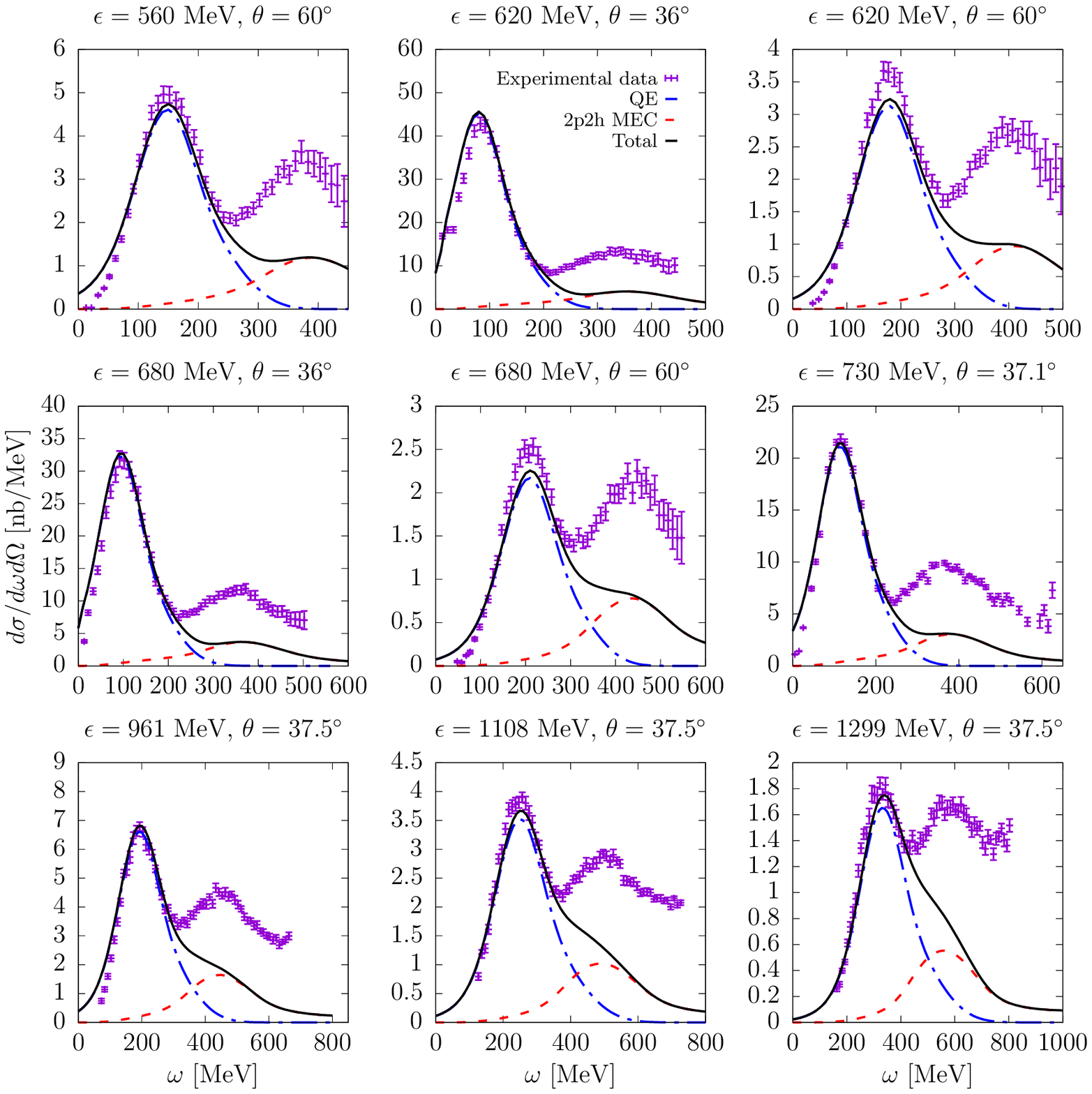}
\caption{ Inclusive electron scattering cross section of $^{12}$C for
  selected kinematics. Results are shown for the new SuSAM* model, with
  the phenomenological scaling function fitted in this work, for the
  2p2h contribution in the RMF model, and the total SuSAM*+MEC*. The
  experimental data data are from ref.  \cite{Ben08,archive,archive2}
}
\label{revalida6}
\end{figure*}

\begin{figure*}
\includegraphics[width=\textwidth, bb=35 440 535 775]{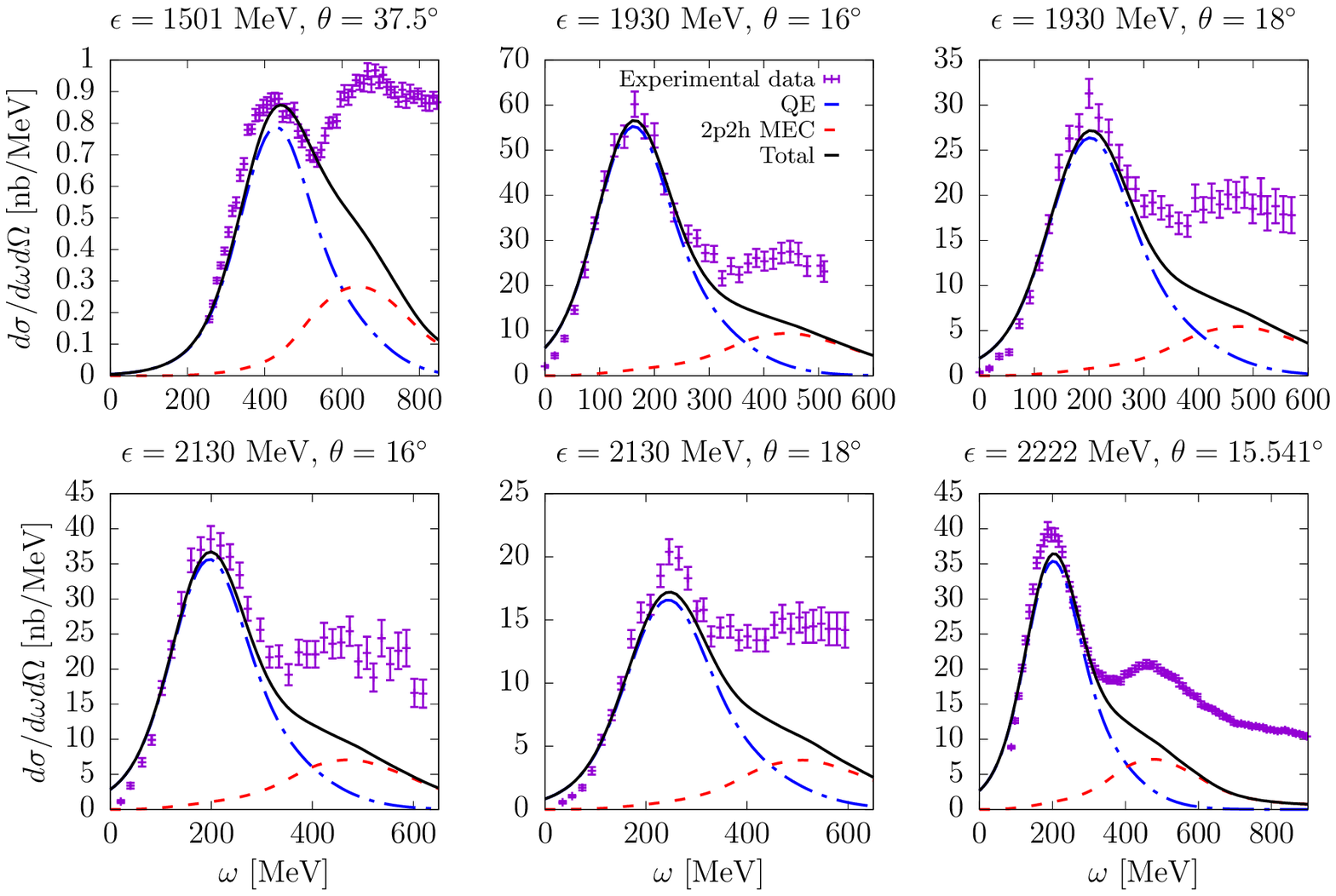}
\caption{
The same as Fig. 8  for other kinematics.
 Data are from ref. 
\cite{Ben08,archive,archive2} 
}
\label{revalida7}
\end{figure*}

In this section we use the phenomenological scaling function obtained
in the previous section to compute the $(e,e')$ cross section, and
evaluate the effect of adding the 2p2h cross section computed in the
RMF.  Since the phenomenological scaling function does not contain
contamination from 2p2h emission, it is safe to add the 2p2h directly
to the SuSAM* model, obtaining a consistent 1p1h+2p2h model with
relativistic effective mass (SuSAM*+MEC*). Pion emission in not
included in the present model. 

Our cross section results are compared to experimental data of
$^{12}$C$(e,e')$ for selected kinematics in Figs. 8 and 9.  In the
last panel of Fig. 9 we also compare with the new data for $^{12}$C
performed in a recent experiment at JLab \cite{Dai18}. In general, the
region of quasielastic peak is well reproduced by the model. The MEC
contribute mainly in the region to the right of the maximum QE peak
(dip region) and go into the pion emission region, where the maximum
of the 2p-2h is reached, contributing to the resonant $\Delta$
peak. The total SuSAM*+MEC* results should be complemented with a pion
emission model in order to reproduce the total cross section in this
region.

\section{Conclusions}

In this work we have extended the superscaling analysis with
relativistic effective mass to include the effects of meson-exchange
currents. The SuSAM* approach takes into account the effects of the
relativistic mean field inside the nucleus, that induces an effective
mass and vector energy to the nucleons.  The effects of 2p2h MEC are
first subtracted from the data before performing the scaling
analysis, and later they are added to the SuSAM* cross section to
obtain the total SuSAM*+MEC* (2p2h) cross section. The MEC matrix
elements are computed in nuclear matter by modifying the nucleon
spinors and the energies according to the solutions of the mean-field
relativistic equation with
scalar and vector potentials in the Walecka model
\cite{Ser86}.  Thus, the 2p2h contribution is
computed using the same ingredients as the SuSAM* 1p1h model, namely
the same value for the relativistic effective mass $m_N^*$. 

A novelty in this work is that the MEC depend on the nucleon vector
energy. That energy does not appear in the 1p1h responses due to the
cancellation between final and initial nucleons.  Using our MEC model
we have been able to estimate the value of the vector energy from the
$(e,e')$ data for $^{12}$C, $E_v \simeq 141$ MeV, and $M^*=0.8$, 
in accord with Serot
and Walecka \cite{Ser86}.

We have verified that the new scaling function $f^*(\psi^*)$ obtained
from the scaling analysis of the $^{12}$C subtracted data
---experimental minus theoretical 2p2h cross section--- is very
similar to the one obtained in a previous work without subtraction of
the MEC contribution. This is because that scaling analysis is based
in a robust data selection method, by elimination of the data that do
not collapse into the quasielastic point cloud. Therefore, in this work we
have shown the strength of the SuSAM* selection method.

Finally, we have computed the total cross section of 1p1h plus 2p2h and
compared to data. The MEC contribution modifies the cross section to
the right of the quasielastic peak, reaching the $\Delta$ peak, 
where the pion emission and inelastic contribution (not included
in this work) are more important.

In future work we will extend this MEC model to the weak sector to
compute the effect of 2p2h in charge-changing neutrino scattering,
that was analyzed with the SuSAM* model, without
including MEC, in Ref. \cite{Rui18}.

\section{Acknowledgments}

This work has been partially supported by the Spanish Ministerio de
Economia y Competitividad (grants No.
FIS2017-85053-C2-1-P) and by the Junta de Andalucia (grant
No. FQM-225).
 V.L.M.C.  acknowledges a contract 
funded by Agencia Estatal de Investigacion and Fondo Social Europeo.



\begin{thebibliography}{99}


\bibitem{Ama20}
J.~E.~Amaro, M.~B.~Barbaro, J.~A.~Caballero, R.~Gonz\'alez-Jim\'enez, G.~D.~Megias and I.~Ruiz Simo,
J. Phys. G \textbf{47} (2020) no.12, 124001

\bibitem{Mos16} U. Mosel, Ann. Rev. Nuc. Part. Sci. 66 (2016), 171.

\bibitem{Kat17}  T.~Katori and M.~Martini,
  J.\ Phys.\ G {\bf 45} (2018) no.1,  013001.

\bibitem{Alv14} 
L. Alvarez-Ruso, Y. Hayato, J. Nieves, New J. Phys. 16 (2014) 075015.

\bibitem{Ank17}  A. M. Ankowski, C. Mariani, J. Phys. G44 (2017) 054001.

\bibitem{Ben17}
 O. Benhar, P. Huber, C. Mariani, D. Meloni, Phys. Rep. 700 (2017) 1.

\bibitem{Nomad09} V Lyubushkin et al.   (NOMAD Collaboration),
 Eur. Phys. J. C 63 (2009), 355.


\bibitem{Agu10} A. Aguilar-Arevalo {\em et al.} (MiniBooNE Collaboration),
Phys. Rev. D {\bf 81}, 092005 (2010).

\bibitem{Agu13} A. Aguilar-Arevalo {\em et al.} (MiniBooNE Collaboration),
Phys. Rev. D {\bf 88}, 032001 (2013).

\bibitem{Fio13} G.A. Fiorentini {\em et al.} (MINERvA Collaboration)
Phys. Rev. Lett. {\bf 111}, 022502 (2013).


\bibitem{Abe13} K. Abe {\em et al.}, (T2K Collaboration),
Phys. Rev. D {\bf 87}, 092003 (2013).

\bibitem{Abe16} K. Abe et al., (T2K Collaboration),
Phys. Rev. D93 (2016) 112012.

\bibitem{Abe18} K. Abe et al. (T2K Collaboration), Phys. Rev. D 97
  (2018), 012001.


\bibitem{Mar09} M. Martini, M. Ericson, G. Chanfray, J. Marteau, 
                 Phys.Rev. C80 (2009) 065501.

\bibitem{Nie11} J. Nieves, I. Ruiz Simo, M.J. Vicente Vacas, 
Phys.Rev. C83 (2011) 045501.

\bibitem{Gal16} 
  K.~Gallmeister, U.~Mosel and J.~Weil,
  Phys.\ Rev.\ C {\bf 94}, no. 3, 035502 (2016).


\bibitem{Meg14} 
  G.~D.~Megias, M.~V.~Ivanov, R.~Gonzalez-Jimenez, M.~B.~Barbaro, J.~A.~Caballero, T.~W.~Donnelly and J.~M.~Udias,
  Phys.\ Rev.\ D {\bf 89}, no. 9, 093002 (2014)
  Erratum: [Phys.\ Rev.\ D {\bf 91}, no. 3, 039903 (2015)]

\bibitem{Ank15} 
  A.~M.~Ankowski,
  Phys.\ Rev.\ D {\bf 92}, no. 1, 013007 (2015).

\bibitem{Gra13} 
  R.~Gran, J.~Nieves, F.~Sanchez and M.~J.~Vicente Vacas,
  Phys.\ Rev.\ D {\bf 88}, no. 11, 113007 (2013).

\bibitem{Pan16} 
  V.~Pandey, N.~Jachowicz, M.~Martini, R.~Gonzalez-Jimenez, J.~Ryckebusch, T.~Van Cuyck and N.~Van Dessel,
  Phys.\ Rev.\ C {\bf 94}, no. 5, 054609 (2016)

\bibitem{Mar16} 
  M.~Martini, N.~Jachowicz, M.~Ericson, V.~Pandey, T.~Van Cuyck and N.~Van Dessel,
  Phys.\ Rev.\ C {\bf 94}, no. 1, 015501 (2016).

\bibitem{Meg16b} 
  G.~D.~Megias, J.~E.~Amaro, M.~B.~Barbaro, J.~A.~Caballero and T.~W.~Donnelly,
  Phys.\ Rev.\ D {\bf 94}, 013012 (2016).

\bibitem{Meg16}
 G.D Megias, 
J.~E.~Amaro, M.~B.~Barbaro, J.~A.~Caballero, T.~W.~Donnelly and I. Ruiz Simo,
 Phys. Rev. D 94 (2016), 093004. 



\bibitem{Lov16}
A. Lovato, S. Gandolfi, J. Carlson, Steven C. Pieper, R. Schiavilla, 
 Phys.Rev.Lett. 117 (2016) 082501.

\bibitem{Pan15} 
  V.~Pandey, N.~Jachowicz, T.~Van Cuyck, J.~Ryckebusch and M.~Martini,
  Phys.\ Rev.\ C {\bf 92}, no. 2, 024606 (2015).

\bibitem{Ank15b} A. M. Ankowski, O. Benhar, M. Sakuda, Phys. Rev. D 91,
  033005 (2015).

\bibitem{Roc16}
N. Rocco, A. Lovato, O. Benhar, 
    Phys.Rev.Lett. 116 (2016)  192501.


\bibitem{Ama02} 
  J.~E.~Amaro, M.~B.~Barbaro, J.~A.~Caballero, T.~W.~Donnelly and A.~Molinari,
  Phys.\ Rept.\  {\bf 368}, 317 (2002).

\bibitem{Ama05} 
  J.~E.~Amaro, M.~B.~Barbaro, J.~A.~Caballero, T.~W.~Donnelly and C.~Maieron,
  Phys.\ Rev.\ C {\bf 71}, 065501 (2005).


\bibitem{Cab07} 
  J.~A.~Caballero, J.~E.~Amaro, M.~B.~Barbaro, T.~W.~Donnelly and J.~M.~Udias,
  Phys.\ Lett.\ B {\bf 653}, 366 (2007).


\bibitem{Udi99} 
  J.~M.~Udias, J.~A.~Caballero, E.~Moya de Guerra, J.~E.~Amaro and T.~W.~Donnelly,
  Phys.\ Rev.\ Lett.\  {\bf 83}, 5451 (1999).

\bibitem{Bod14} A. Bodek, M.E. Christy, and B. Coopersmith, 
Eur. Phys. Jou. C 74, 3091 (2014). 


\bibitem{Gil97}
A.~Gil, J.~Nieves and E.~Oset,
Nucl.\ Phys.\ A  {\bf 627} (1997) 543.

\bibitem{Sim16} 
  I.~Ruiz Simo, J.~E.~Amaro, M.~B.~Barbaro, A.~De Pace, J.~A.~Caballero and T.~W.~Donnelly,
J.Phys. G44 (2017) no.6, 065105.



\bibitem{Nie17} J. Nieves, J.E. Sobczyk,   Annals Phys. 383 (2017) 455.

\bibitem{Alb88} W.M. Alberico, A. Molinari, T.W. Donnelly, E. L. Kronenberg, 
and J.W. Van Orden, Phys Rev. C 38 (1988) 1801.

\bibitem{Don99a} 
  T.~W.~Donnelly and I.~Sick,
  Phys.\ Rev.\ Lett.\  {\bf 82}, 3212 (1999).

\bibitem{Don99b} 
  T.~W.~Donnelly and I.~Sick,
  Phys.\ Rev.\ C {\bf 60}, 065502 (1999).


\bibitem{Mai02} 
  C.~Maieron, T.~W.~Donnelly and I.~Sick,
  Phys.\ Rev.\ C {\bf 65}, 025502 (2002).

\bibitem{Ama04} J.~E.~Amaro, M.~B.~Barbaro, J.~A.~Caballero,
                T.~W.~Donnelly, A. Molinari, I. Sick, 
   Phys. Rev. C {\bf 71}, 015501  (2005).

\bibitem{Hor91} C.J. Horowitz, D.P. Murdock, and B.D. Serot,
in {\em Computational Nuclear Physics Vol. 1}, Springer-Verlag, Berlin 1991.

\bibitem{Gon14} R. Gonzalez-Jimenez, G.D. Megias, M.B. Barbaro, J.A. Caballero, 
and T.W. Donnelly, Phys. Rev. C 90, 035501 (2014).



\bibitem{Ama15} 
  J.~E.~Amaro, E.~Ruiz Arriola and I.~Ruiz Simo,
  Phys.\ Rev.\ C {\bf 92}, no. 5, 054607 (2015).

\bibitem{Ama17} J.~E.~Amaro and E.~Ruiz Arriola,    I.~Ruiz Simo, 
  Phys.\ Rev.\ D {\bf 95}, 076009 (2017).

\bibitem{Mar17} 
  V.~L.~Martinez-Consentino, I.~Ruiz Simo, J.~E.~Amaro and E.~Ruiz Arriola,
  Phys.\ Rev.\ C {\bf 96}, no. 6, 064612 (2017).

\bibitem{Ama18}
J.~E.~Amaro, V.~L.~Martinez-Consentino, E.~Ruiz Arriola and I.~Ruiz Simo,
Phys. Rev. C \textbf{98} (2018) 024627



\bibitem{Rui18} I.~Ruiz Simo,   V.~L.~Martinez-Consentino, 
                J.~E.~Amaro and E.~Ruiz Arriola,
                Phys.\ Rev.\ D {\bf 97}, 116006 (2018).


\bibitem{Ros80} R. Rosenfelder, Ann. Phys. (N.Y.) 128, 188 (1980).

\bibitem{Ser86} B.D. Serot, and J.D. Walecka, Adv. Nucl. Phys. 16 (1986) 1.

\bibitem{Weh93} K. Wehrberger, Phys. Rep. 225 (1993) 273.

\bibitem{archive} O. Benhar, D. Day and I. Sick, arXiv:nucl-ex/0603032.

\bibitem{archive2} O. Benhar, D. Day, and I. Sick, 
http://faculty.virginia.edu/qes-archive/  

\bibitem{Ben08} O. Benhar, D. Day, and I. Sick, Rev Mod Phys. 80 (2008) 189.

\bibitem{Dai18} H. Dai {\em et al.,} (JLab Hall A Collaboration), 
arXiv:1803.01910 [nucl-ex]. 


\bibitem{Bar98} M.B. Barbaro, R. Cenni, A. De Pace, T.W. Donnelly,
  A. Molinari, Nucl. Phys. A 643 (1998) 137.


\bibitem{For83} 
  T.~De Forest,
  Nucl.\ Phys.\ A {\bf 392}, 232 (1983).

\bibitem{Gal71} 
  S.~Galster, H.~Klein, J.~Moritz, K.~H.~Schmidt, D.~Wegener and J.~Bleckwenn,
  Nucl.\ Phys.\ B {\bf 32}, 221 (1971).


\bibitem{Sim14}
I.~Ruiz Simo, C.~Albertus, J.~E.~Amaro, M.~B.~Barbaro, J.~A.~Caballero and T.~W.~Donnelly,
Phys. Rev. D \textbf{90} (2014) no.3, 033012

\bibitem{DeP03}
A.~De Pace, M.~Nardi, W.~M.~Alberico, T.~W.~Donnelly and A.~Molinari,
Nucl. Phys. A \textbf{726} (2003), 303-326

\bibitem{Som78} B. Sommer, Nucl. Phys A 308 (1978) 263.

\bibitem{Alb84} W. Alberico, M. Ericson, and A. Molinari, Ann. Phys. (N.Y.) 154 (1984) 356.

\bibitem{Dek94} M. J. Dekker, P. J. Brussaard, and J. A. Tjon
Phys. Rev. C 49 (1994) 2650

\bibitem{Hernandez:2007qq}
  E.~Hernandez, J.~Nieves and M.~Valverde,
  Phys.\ Rev.\ D {\bf 76} (2007) 033005.



\bibitem{Sim17}
I.~Ruiz Simo, J.~E.~Amaro, M.~B.~Barbaro, J.~A.~Caballero, G.~D.~Megias and T.~W.~Donnelly,
Phys. Lett. B \textbf{770}, 193-199 (2017)



\end{thebibliography}
\end{document}